\documentclass[runningheads]{llncs}
\usepackage[T1]{fontenc}
\usepackage[utf8]{inputenc}
\usepackage[russian,english]{babel}
\usepackage{graphicx}

\usepackage{hyperref}
\usepackage{todonotes}
\usepackage{color}

\usepackage{multirow}
\usepackage{lscape}
\usepackage{pifont}
\usepackage{numprint}

\npthousandsep{`} 
\npthousandthpartsep{}
\npdecimalsign{.}

\begin{document}

\title{You were saying? --\\ Spoken Language in the V3C Dataset}

\author{Luca Rossetto}

\institute{
Department of Informatics\\
University of Zurich\\
Zurich, Switzerland\\
\email{rossetto@ifi.uzh.ch}
}

\maketitle

\begin{abstract}

This paper presents an analysis of the distribution of spoken language in the V3C video retrieval benchmark dataset based on automatically generated transcripts. It finds that a large portion of the dataset is covered by spoken language. Since language transcripts can be quickly and accurately described, this has implications for retrieval tasks such as known-item search.


\end{abstract}

\section{Introduction}

Spoken language is a salient component of video that is often not only easy to remember but one that can also be quickly and accurately reproduced. This is especially relevant for known-item search scenarios, where previously encountered video content is to be retrieved. While many different query modes can be found in content-based video retrieval, expressing one's information need in textual form is still among the most intuitive and poses few requirements for the user.

The Vimeo Creative Commons Collection (V3C)~\cite{v3c} is a video dataset used in retrieval benchmarks such as the Video Browser Showdown~\cite{DBLP:journals/ijmir/HellerGBG0LLMPR22} or TRECVID~\cite{2022trecvidawad}. It is designed to be generally representative of web video as a whole and is distributed in three shards: V3C1, V3C2, and V3C3.

This paper presents an analysis of automatically generated transcriptions of V3C and discusses some of its findings as well as their consequences for video retrieval. It looks at the languages detected in the dataset and the relative amount of playback time they cover. Section~\ref{sec:method} outlines the process used for the analysis, and Section~\ref{sec:results} presents its results. Finally, Section~\ref{sec:conclusion} offers some discussion and concluding remarks. The data and code used for the analysis are available via \url{https://github.com/lucaro/V3C-language-analysis}.

\pagebreak

\section{Method}
\label{sec:method}
This analysis is based on previously released~\cite{transcripts_zenodo} transcripts generated using \mbox{OpenAI's} \emph{Whisper}~\cite{whisper} model, specifically the `small' variant. These transcripts come in the form of subtitles which contain a series of textual elements associated with a time interval during which this text was detected to be spoken. Despite the high transcription quality offered by the Whisper model, applying it to unconstrained data such as the V3C produces output that is noisy in some cases. Therefore, the first step of the analysis process required some data cleaning in order to improve the overall data quality.

\subsection{Data-Cleaning}

While the transcription quality is reasonably high during parts of the video with clearly audible speech, other types of audio content can produce various transcription artifacts. In order to remove as many of these as possible for downstream analysis, simple filtering heuristics are applied. This filtering is performed on the basis of individual subtitle elements. Elements that contain fewer than 6 characters or that span a time interval of fewer than 0.5 seconds are discarded since they are unlikely to include enough information for subsequent analysis. Additionally, all elements that primarily consist of white space or punctuation marks are removed. This filtering step cuts down on subtitle elements erroneously generated by non-speech audio, such as noise. There is, however, one particular type of non-speech sound that commonly occurs in video: background music. The next step in the pipeline deals with this content specifically.

\subsection{Music Estimation}

A common non-speech component of audio found in video is music. While Whisper was not explicitly trained to detect music, some distinct types of transcriptions can be observed whenever music is playing. This is presumably due to some examples in the training set where musical interludes were annotated with subtitles. Most commonly, whisper will just generate the transcript \emph{``Music''} during that time, but several other outputs can be observed as well. Apparently, the training set contained several Russian language examples where musical passages were transcribed with different phrases describing the type of music playing. This can be deduced from the transcription output, that commonly also contains terms such as \emph{``\foreignlanguage{russian}{интригующая музыка}'' (intriguing music)}, \emph{``\foreignlanguage{russian}{позитивающая музыка}'' (positive music)}, etc. While no comprehensive list of such transcription outputs can be generated, random sampling of video instances with music yields a short list of commonly occurring terms that can be used heuristically for music estimation. When a subtitle element primarily consists of such terms, they are not treated as transcriptions of spoken language but rather as referring to music playing at this point in time.

\begin{figure}[t]
    \centering
    \includegraphics{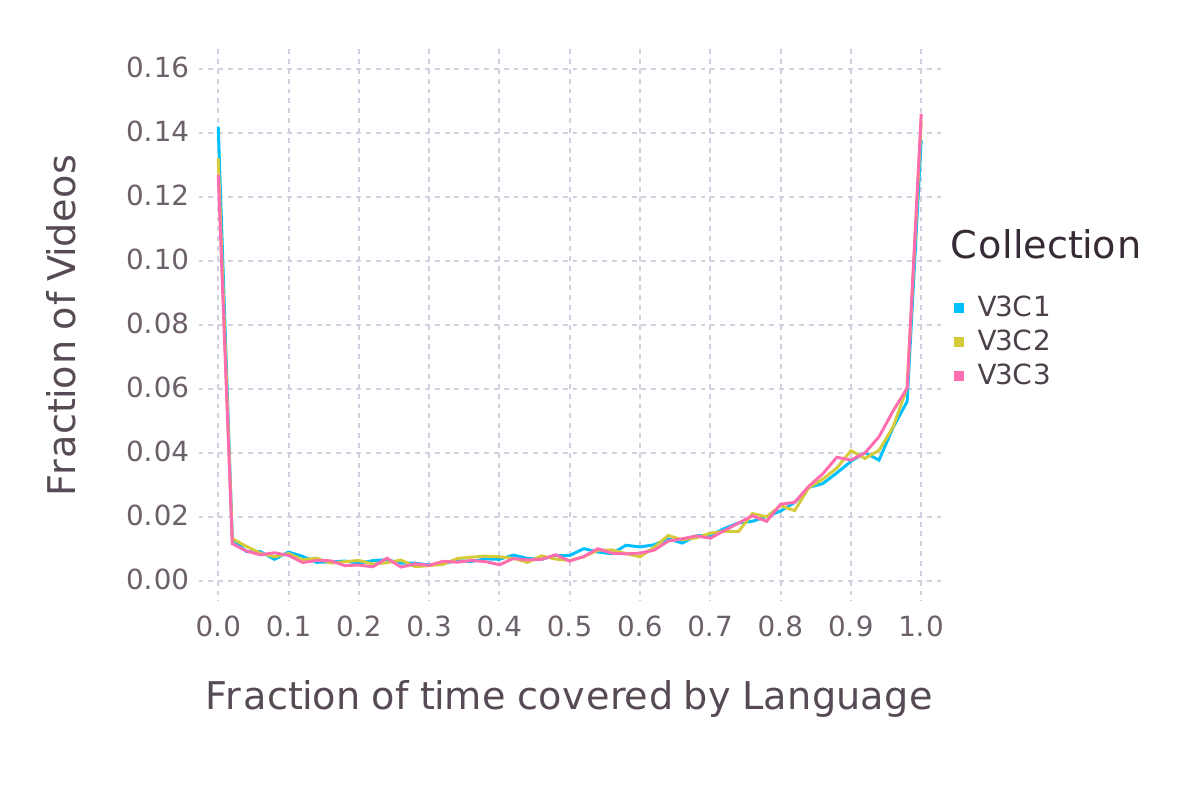}
    \caption{Distribution of fraction of video duration during which transcribed language is present.}
    \label{fig:language_fraction}
\end{figure}

\subsection{Language Detection}
\label{sec:language-detection}
For all subtitle elements that passed the relevance threshold and were not considered to refer to music, language detection was performed using the \emph{Lingua}\footnote{\url{https://github.com/pemistahl/lingua}} natural language detection library in version 1.2.2. This library is capable of identifying 75 different languages based on even short strings, which makes it well-suited to the application of individual subtitle elements. In cases where \emph{Lingua} was unable to identify a language, every character in the string was mapped to its corresponding \emph{Unicode Script}\footnote{\url{https://www.unicode.org/Public/UNIDATA/Scripts.txt}} and the most common one was used as a stand-in for the language.

\section{Results}
\label{sec:results}

Across the entire V3C dataset, language was transcribed for 68.95\% of the cumulative video time. Put differently, when selecting a video and point in time at random, one has an almost 70\% chance of there being words spoken at that time. This is, however, not uniform across all videos, as can be seen in Figure~\ref{fig:language_fraction}. The figure shows two clear outliers at 0\% and 100\% relative coverage of transcripts. Of the videos that contain any transcripts but are not completely covered by them, the figure still shows a clear skew towards more language coverage rather than less. Roughly 4\% of all videos have transcripts for 90\% of their duration, and about 2\% of videos are covered to 80\% by transcripts.

\begin{figure}[t]
    \centering
    \includegraphics{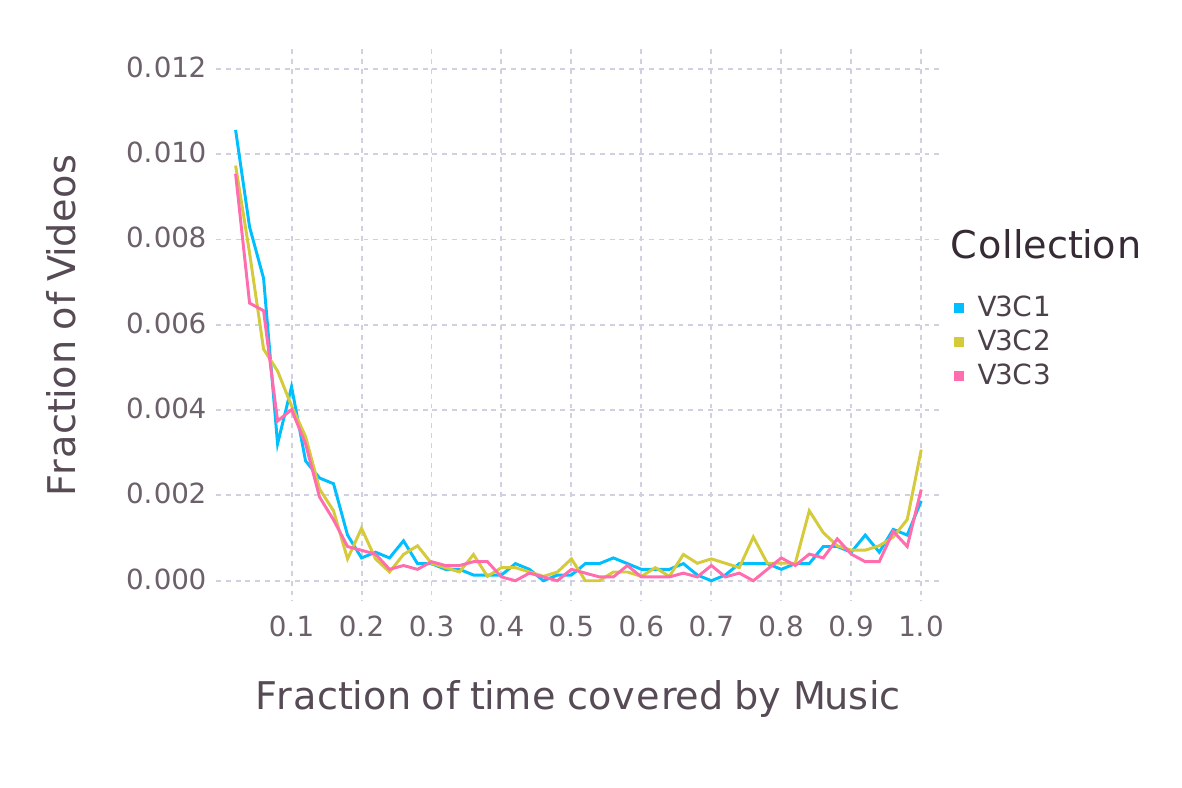}
    \caption{Distribution of fraction of video duration during which music was estimated to be present. Only showing videos where any music was estimated.}
    \label{fig:music_fraction}
\end{figure}

The prevalence of music, as estimated from the transcripts, is much lower. Only about 1.36\% of total video playback time is covered by transcripts that were estimated to refer to music. This is almost certainly an underestimation, as only time sequences that are exclusively filled with music will be transcribed as such. If there is clearly audible speech on top of music, including singing, the model tends to produce transcripts of the spoken words without any indication of background music. Figure~\ref{fig:music_fraction} shows the relative coverage of all the videos for which any music was detected. It can be seen that only a tiny fraction of videos has a high music coverage. This is to be expected since `music videos' that were explicitly labeled as such were deliberately excluded during the dataset's compilation.

Based on the pipeline outlined in Section~\ref{sec:language-detection}, Table~\ref{tab:time-coverage} shows the top 10 identified languages sorted by the amount of time their transcripts cover, as well as the amount of estimated music and silence. Spoken words in English are in the clear majority, covering 45.4\% of the playback time across the entire dataset. Since 29.7\% of playback time has no transcripts and 1.4\% are exclusively covered by music, spoken English occupies more time in the dataset than all other languages combined. The next 5 most common languages are Spanish, French, Portuguese, Italian, and German, with Russian coming in at rank 7 and covering less time than music. This means that the 6 most commonly occurring spoken languages in the dataset, collectively covering 59.3\% of playback time, all use the Latin alphabet and can (mostly) be expressed using ASCII characters.

\begin{table}[t]
\centering
\caption{Top 10 detected languages with respect to the amount of time their transcripts cover (including no transcripts and estimated music) per dataset shard.}
\label{tab:time-coverage}
\def\arraystretch{1.3}
\begin{tabular}{l|c|c|c|c|}
\cline{2-5}
                                     & V3C1   & V3C2   & V3C3   & V3C (total) \\ \hline
\multicolumn{1}{|l|}{English}        & 45.4\% & 44.4\% & 46.2\% & 45.4\%      \\ \hline
\multicolumn{1}{|l|}{\textit{None}}  & 30.5\% & 30.4\% & 28.5\% & 29.7\%      \\ \hline
\multicolumn{1}{|l|}{Spanish}        & 4.3\%  & 4.5\%  & 4.4\%  & 4.4\%       \\ \hline
\multicolumn{1}{|l|}{French}         & 3.9\%  & 4.2\%  & 4.2\%  & 4.1\%       \\ \hline
\multicolumn{1}{|l|}{Portuguese}     & 1.7\%  & 2.0\%  & 2.1\%  & 1.9\%       \\ \hline
\multicolumn{1}{|l|}{Italian}        & 1.6\%  & 1.4\%  & 2.0\%  & 1.7\%       \\ \hline
\multicolumn{1}{|l|}{German}         & 1.7\%  & 1.2\%  & 2.0\%  & 1.7\%       \\ \hline
\multicolumn{1}{|l|}{\textit{Music}} & 1.3\%  & 1.8\%  & 1.1\%  & 1.4\%       \\ \hline
\multicolumn{1}{|l|}{Russian}        & 0.7\%  & 1.1\%  & 0.7\%  & 0.9\%       \\ \hline
\multicolumn{1}{|l|}{Sinhalese}      & 0.7\%  & 0.6\%  & 0.5\%  & 0.6\%       \\ \hline
\multicolumn{1}{|l|}{Bulgarian}      & 0.7\%  & 0.5\%  & 0.4\%  & 0.5\%       \\ \hline
\multicolumn{1}{|l|}{Dutch}          & 0.5\%  & 0.4\%  & 0.6\%  & 0.5\%       \\ \hline
\end{tabular}
\end{table}

\section{Conclusion}
\label{sec:conclusion}

In the context of video retrieval, spoken language is an effective query mode, as words can be used for query formulation accurately, easily, and losslessly. The analysis of automatically generated transcripts generated from the Vimeo Creative Commons Collection (V3C) indicates that spoken language is sufficiently common in web video to make this approach effective. When picking a point in time from any video in the dataset at random, there is a greater than 45\% chance that there will be spoken English and a roughly 60\% chance of encountering a spoken language that can be written (at least approximately) using only ASCII characters. 
While automatic speech transcription has yet to reach parity with human performance, the output generated by modern methods is sufficiently accurate for many retrieval applications, even when applied to unconstrained video content.

\section*{Acknowledgements}
This work was partly supported by the Swiss National Science Foundation through project \href{https://data.snf.ch/grants/grant/202125}{``MediaGraph''} (contract no.\ 202125).

\bibliographystyle{splncs04}
\bibliography{bibliography}

\end{document}